\documentclass[printer]{aa}
\usepackage[varg]{txfonts}

\usepackage{graphicx}
\usepackage{natbib}
\usepackage{color}
\usepackage{subfig}
\usepackage[switch]{lineno} 
\usepackage{xspace}
\usepackage{microtype}
\usepackage{amssymb}
\def\link_col{blue}
\usepackage{xspace}
\usepackage{url}
\usepackage{wasysym}
\usepackage{caption}
\usepackage[normalem]{ulem}

\def\grays{$\gamma$-rays\xspace}
\def\gray{$\gamma$-ray\xspace}

\begin{document}
\title{Diffuse \gray emission in the vicinity of  young star cluster Westerlund 2}
\titlerunning{\grays from Westerlund 2}
\author{Rui-zhi Yang\inst{1}
\and Emma de O\~na Wilhelmi\inst{2}
\and Felix Aharonian\inst{1, 3}}
\institute{Max-Planck-Institut f{\"u}r Kernphysik, P.O. Box 103980, 69029 Heidelberg, Germany.
\and Institute for Space Sciences (CSIC$--$IEEC), Campus UAB, Carrer de Can Magrans s/n, 08193 Barcelona, Spain.
\and Dublin Institute for Advanced Studies, 31 Fitzwilliam Place, Dublin 2, Ireland.
}%
\date{Received:  / Accepted: } 
\abstract
{We report  the results of  our analysis of  the publicly available  data obtained by the  Large Area Telescope (LAT) on board of the {\it Fermi} satellite towards the direction of the young massive star cluster Westerlund 2. We found significant extended \gray emission in the vicinity of Westerlund 2 with a hard power-law energy spectrum extending from 1 GeV to 250 GeV  with a photon index of $2.0 \pm 0.1$. We argue that amongst  several alternatives, the luminous  stars in Westerlund 2 are likely sites of  acceleration of particles responsible for the diffuse $\gamma$-ray emission of the surrounding interstellar medium.  In particular, the young star cluster Westerlund 2 can provide sufficient non-thermal energy  to account for the gamma-ray emission.  In this scenario, since the \gray production region is significantly larger than the area occupied by the star  cluster, we conclude that the \gray production is caused by  
hadronic interactions of the accelerated protons and nuclei with the ambient gas. In that case, the total  energy budget in relativistic particles  is  estimated of the order of $10^{50} ~\rm erg$.}
\maketitle
\section{Introduction}
Westerlund~2 is one of the most massive young star clusters in our Galaxy \citep[see, e.g., ][]{zwart10}. It has an age  of (2 - 3) $ \times 10^6\rm $ yrs \citep{piatti98} and a total stellar mass of about (1 - 3) $\times 10^4\rm~ M_{\odot}$ \citep{ascenso07}.   It hosts at least a dozen of O stars and two remarkable Wolf-Rayet (WR) stars, WR~20a and WR~20b, and causes ionisation of  the nearby  \ion{H}{ii} region RCW~49. Radio continuum observations reveal two wind-blown shells in the core of RCW~49 \citep{whiteoak97} surrounding the central region of Westerlund~2 and  the the star WR 20b. Two molecular clouds are also observed in this region, whose velocity dispersion reveals a hint of the cloud collision. The distance to Westerlund~2 is estimated to be $5.4^{+ 1.1}_{- 1.4}$ kpc \citep{furukawa09} or $8.0 \pm 1.4$ kpc \citep{rauw07}. 

Westerlund 2 has been observed at very high energy (VHE) \grays with the H.E.S.S. array of imaging atmospheric Cherenkov telescopes \citep{hess_w2, hess_w2old}. Two extended VHE \gray sources,
HESS~J1023--575 and HESS~J1026--582, have been detected in this region, the origins of which are still under debate. They may come from the pulsar wind nebulae of the powerful pulsars PSR~J1022--5746 and PSR~J1028--5819  or from the collective stellar wind interaction in the young clusters.  Recently, \citet{fges} reported the detection of two extended sources emitting in the GeV range in the region: FGES~J1023.3--5747 is spatially coincident with the TeV source HESS~J1023--575; the second source, FGES~J1036.4--5834, is a very extended source with a radius of more than $2^{\rm o}$ and a hard spectrum with a photon index of about 2. The large spatial extension and the hard spectrum  remind  the \gray emission from other young star clusters or superbubble systems such as Cygnus cocoon \citep{fermi_cygnus} or NGC 3603 \citep{yang17}. 

In this paper we investigate this scenario by analysing LAT data obtained towards the direction of Westerlund 2, with a focus on the very extended  component of $\gamma$-rays, and discuss  the possible origin of this radiation.

\section{Fermi LAT data analysis}
\subsection{Spatial analysis}
We selected LAT data  towards Westerlund 2  for a period of approximately 7 years (MET 239557417 -- MET 455067824).
For the analysis, we have used the standard LAT analysis software package \emph{v10r0p5}\footnote{\url{http://fermi.gsfc.nasa.gov/ssc}}. 

The region-of-interest (ROI) was selected to be a $15^{\circ} \times 15^{\circ}$ square centred on the position of Westerlund 2, i.e., $RA_{\rm J2000}=155.992^{\circ}$, $DEC_{\rm J2000}=-57.764^{\circ}$. Observations with Rock angle larger than $52^{\circ}$ are excluded in this analysis. 
In order to reduce the effect of the Earth albedo background, we excluded  the time intervals when 
the parts of the ROI were observed at zenith angles $> 90^{\circ}$. %
 For the spatial analysis, given the crowded nature of the region and to avoid systematic errors due to a poorer angular resolution at low energies, we selected only the photons with energies exceeding 10~GeV. This energy cut reduces the contribution of bright, nearby pulsars, allowing to describe the emission with an angular resolution of $\sim0.1^{\circ}$.
The   $P8\_R2\_v6$ version of the  post-launch instrument response functions (IRFs) was used, and both the front and back converted photons were selected.

\begin{table}[h]
\centering
\footnotesize
\caption{Fitting results for the different spatial models explained in the text. All of them include the diffuse component.}
\label{tab:loglike}
\begin{tabular}{clclclc}
\hline
\hline
 & -$\log({\cal L})$ &  TS & d.o.f \\
\hline
Model 1 & 30633 & & 83\\
Model 2 & 30390 & 486& 87\\
Model 3   &30577 & &122\\
Model 4 & 30683 & & 77\\
Model 5 & 30355 & 486 + 70& 87\\
\hline
\hline
\end{tabular}
\end{table}
\begin{figure*}
\centering
\includegraphics[width=\linewidth]{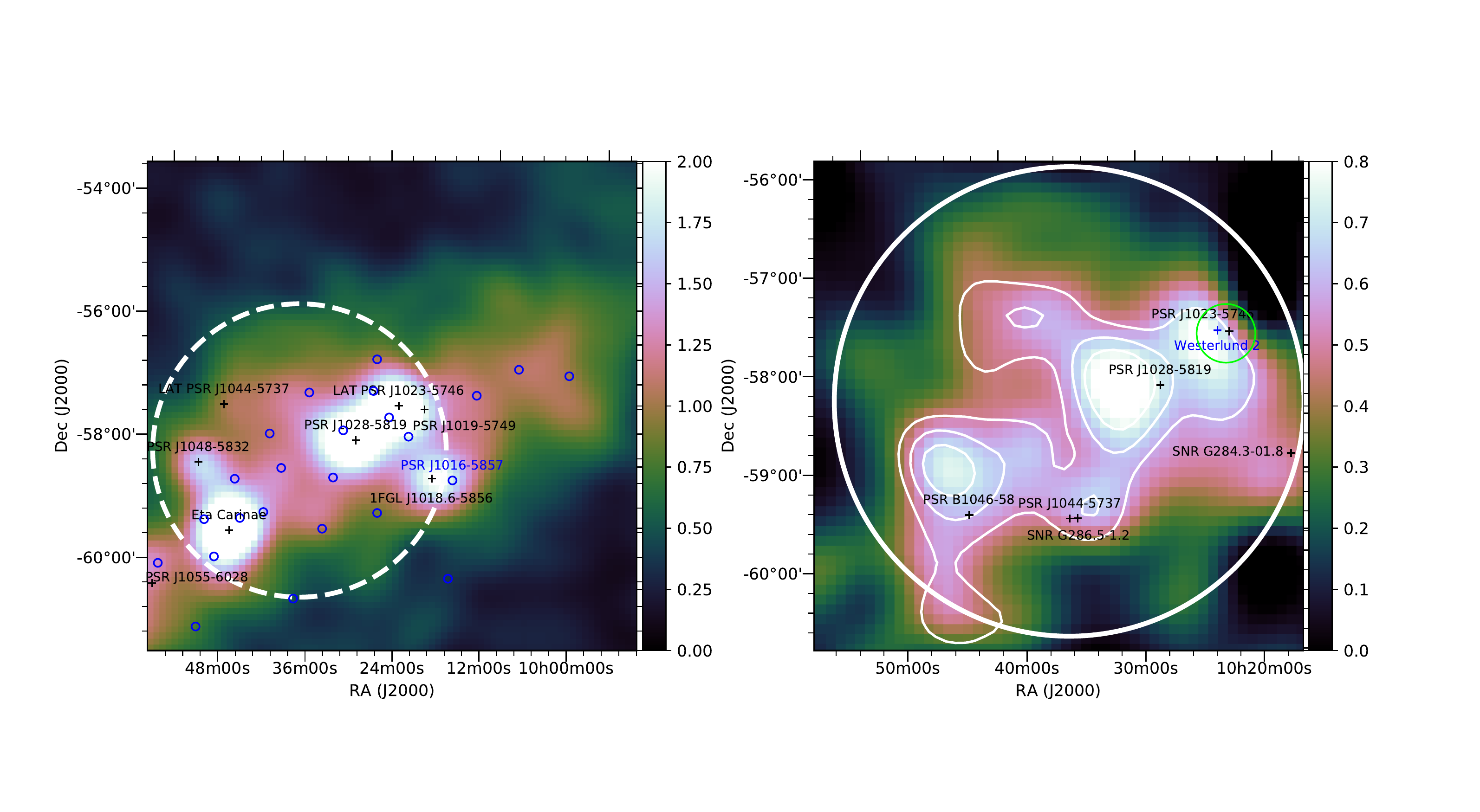}
\caption{On the left a) \grays counts map above 10~GeV in the $8^{\circ} \times 8^{\circ}$ region around Westerlund 2. The 3FGL catalog sources are
labeled with a black cross and blue circle for identified and unassociated sources, respectively. The white dashed line shows the
extended emission related to FGES~J1036.4--5834. On the right b) Residual map after subtracting all the identified catalog source
and the diffuse background, centred on the best-fit position of the extended source to the south of Westerlund 2. Also shown are the
best-fit 2D disk template (white circle), the position and extension of FGES~J1023--5746 (black circle), and the known SNRs and
bright pulsars in the field of view. The extension of the H.E.S.S. source HESS~J1026--582 is also marked with a black circle. The
white contours are obtained from the TS-map, for levels 3, 4 and 5.
}
\label{fig:cmap}
\end{figure*}

The \gray count map above 10 GeV  in  the  $8^{\circ} \times 8^{\circ} $ region around Westerlund 2 is shown on the  left panel of  Fig.~\ref{fig:cmap} (a). We performed a binned likelihood analysis by using the tool {\it gtlike}.  The point sources listed in the 3rd Fermi source catalog (3FGL) \citep{3fgl} are also shown with a black cross, in case of identified ones, and with a blue circle in case of non-associated ones. Those non-associated point-like sources could be part of the extended emission, therefore, in the likelihood fitting, within two degrees from Westerlund 2, we first included all  sources from the 3FGL catalog, but we excluded the ones without an association to a known source (Model 1). We also added  the background models provided by the Fermi collaboration (gll\_iem\_v06.fits and  \_P8R2\_SOURCE\_V6\_v06.txt  for the galactic and the isotropic diffuse  components, respectively\footnote{ available 
at \url{http://fermi.gsfc.nasa.gov/ssc/data/access/lat/BackgroundModels.html}}).
  In the analysis, the normalisations and the spectral indices of sources in side the FOV were left free.  

After the likelihood fitting (-$\log({\cal L})$), we subtracted the best-fit diffuse model and all the identified point sources in the ROI. The resulted residual map is shown in the right panel of Fig.~\ref{fig:cmap} (b). We found substantial residual emission within a large spatial extension. We modelled those as a uniform disk (Model 2) and then varied the position and the size of the disk to find the best fit parameters.  The best-fit result is a uniform disk centred at ($RA_{\rm J2000}=(159.10\pm 0.10)^{\circ}$,  $DEC_{\rm J2000}=(-58.50\pm 0.10)^{\circ}$) with a radius of $r=2.4^{+0.1~\circ}_{-0.2}$. The TS-value is about 500, corresponding to a statistical significance of more than 22$\sigma$.  

We also tested whether this extended emission is composed by several independent point-like sources.  To do this, we included all sources listed in the 3FGL catalogue (Model 3). The resulted -$\log({\cal L})$ function value is larger than that in the extended template case, even with much more free parameters. To account for the fact that we are selecting only high energy events (10~GeV), we also tested a model including all the sources in 3FHL \citep {3fhl} catalog instead of 3FGL catalog (Model 4). No improvement is obtained in the global statistics. Finally, we added FGES~J1023--5746 as a disk to Model 2 \citep{fges} with a radius of $0.28^{\circ}$ in the model (referred as Model 5 in Table \ref{tab:loglike}) and obtain a TS-value for this extended source of 70. Therefore, in the following analysis we included both the $2.4^{\circ}$ disk and  $0.28^{\circ}$ disk as  spatial templates.   We use the position and extension of  FGES~J1023--5746  derived by  \citet{fges}. The results described above for the different models are listed in Table \ref{tab:loglike}. 
 
\subsection{Spectral analysis}

For the spectral analysis  we  applied  \emph{gtlike} in the  energy range [0.3, 250] GeV and modelled the spectrum of the $2.4^{\circ}$ disk as a power-law function, fixing the position to the one found in the spatial analysis.  We neglected the emission below 0.3 GeV to avoid systematic errors due to the dominant Galactic diffuse background. The derived  photon index is 
 $2.02 \pm 0.11 (stat) \pm 0.1(sys)$, and  the total flux above 1 GeV is $4.2 \pm 0.2 \times 10^{-8} ~\rm cm^{-2} s^{-1}$. The systematic errors come from the uncertainties of the effective area and the point spread function of LAT \citep{fermi_pass7}.  The derived spectrum is in consistency with the results of FGES~J1036.4--5834 in \citet{fges}, which is $2.08 \pm 0.06 \pm 0.07$. 

\begin{figure}
\centering
\includegraphics[width=\linewidth]{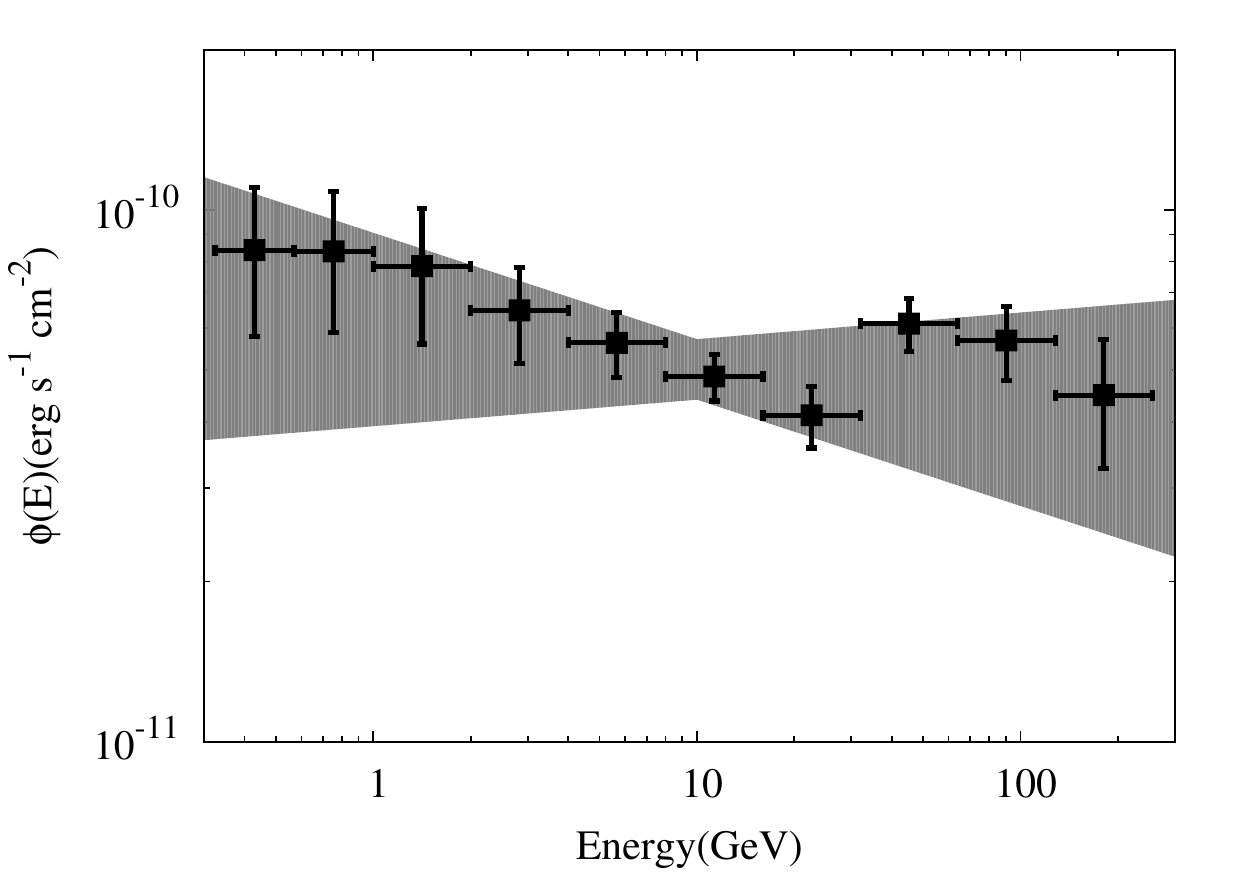}
\caption{The SEDs of the extended emission of the $2.4^{\circ}$ disk. The grey region represents the statistical error when fitting the global spectrum with a power-law function. 
}
\label{fig:SED}
\end{figure}

To obtain the spectral energy distribution (SED) of extended emission to the South of Westerlund 2, we divided the full energy range into 10 logarithmically spaced bands and applied \emph{gtlike} to each of these bands, in this process the index was fixed to 2. The results of this analysis are shown in Fig.~\ref{fig:SED}. All data points have TS values larger than 4, which corresponds to a significance of larger than $\sim2\sigma$.  The power-law spectrum extends to 250 GeV without any sign of a cutoff or a break.  

To test the impact of uncertainties characterising  the diffuse Galactic background models, 
 in the likelihood fitting we artificially varied the galactic diffuse background  by 6 percent to 
 estimate the influence on the final flux.  While below 1 GeV the uncertainties due to the varied diffuse background can be as large as 30~\%, they become  negligible above 4 GeV. These uncertainties are  included in the error bars shown in Fig.~\ref{fig:SED}. 

\subsection{Pulsar gating analysis on PSR~J1022--5746}

\begin{figure*}
\centering
\includegraphics[width=0.45\linewidth]{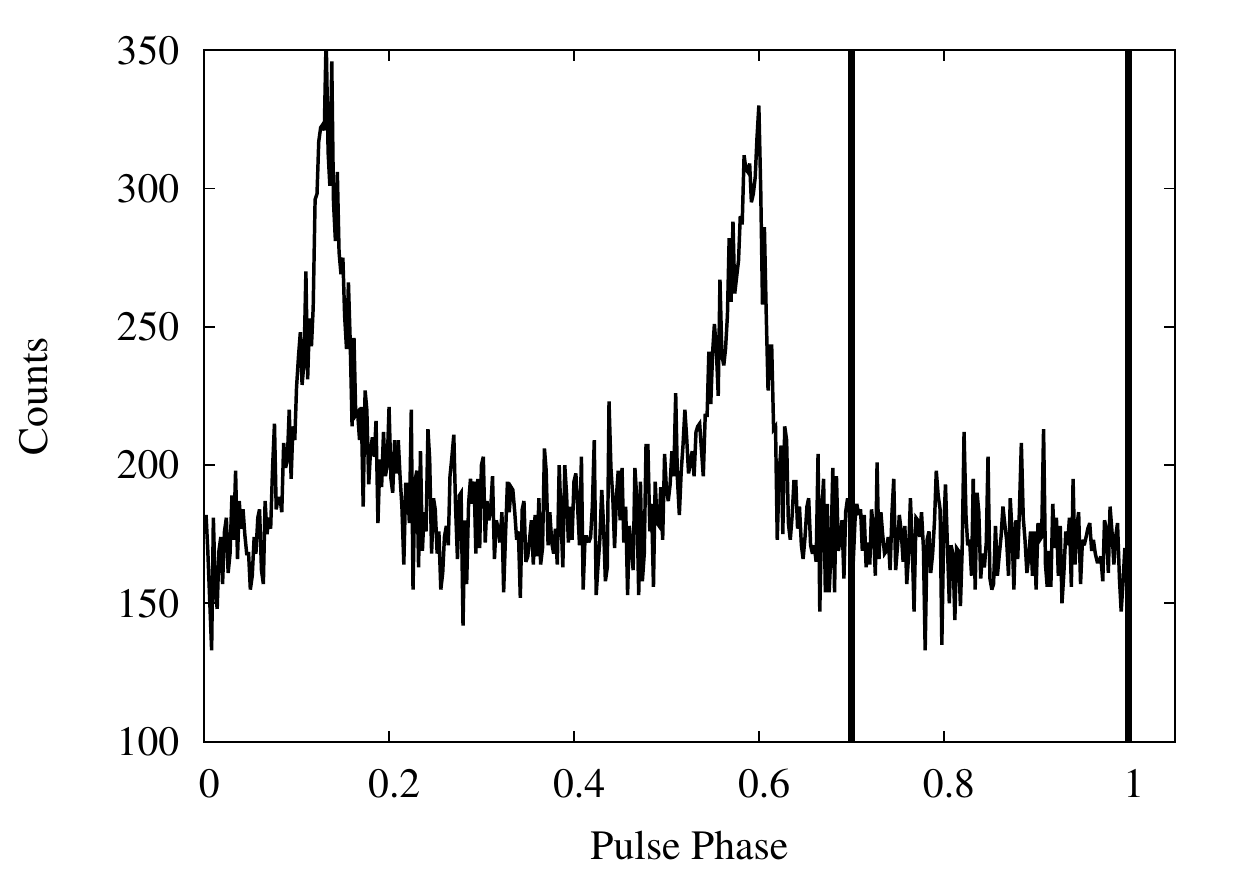}
\includegraphics[width=0.45\linewidth]{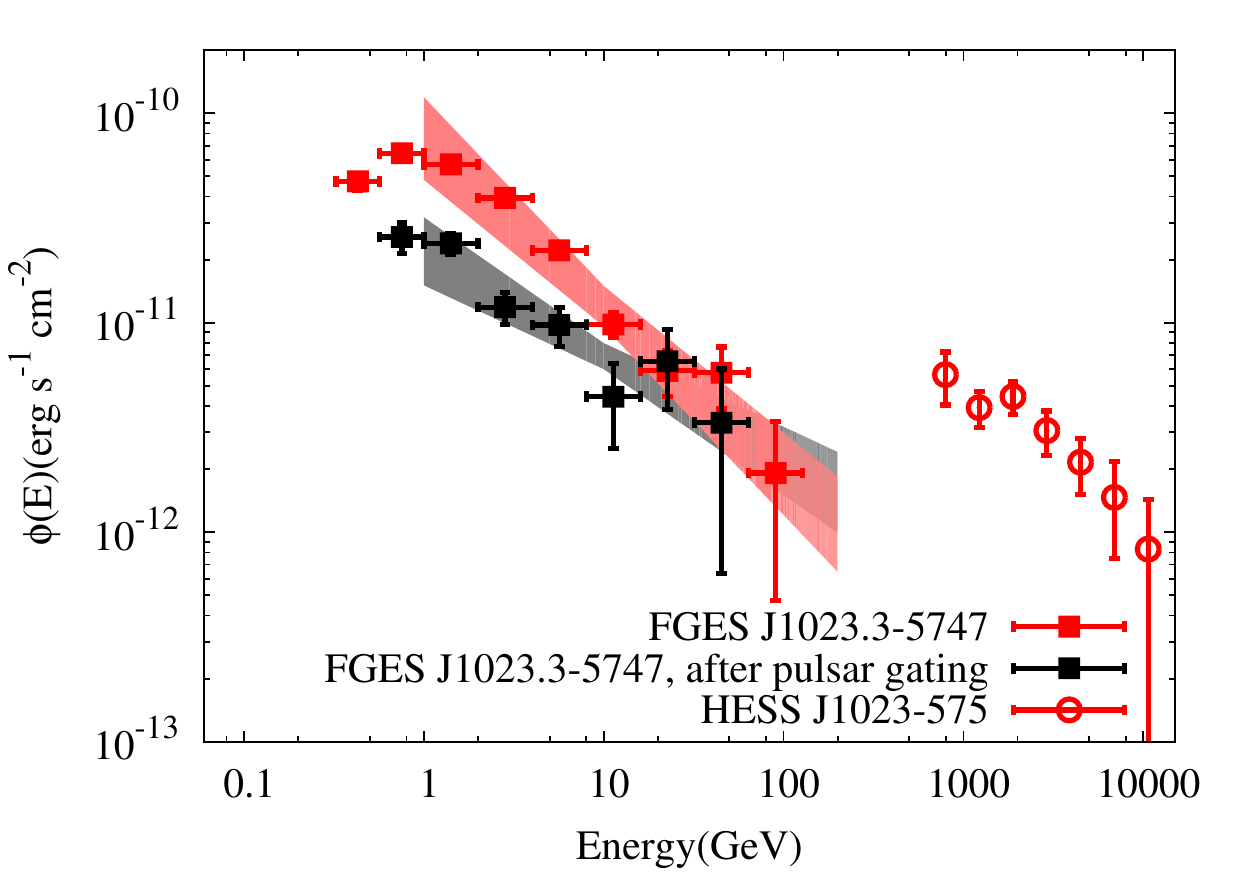}
\caption{On the left a) The phase-folded light curve of pulsar PSR J1022--5746. The two vertical lines indicate the interval in which we selected in the pulsar gating analysis. on the right b) SEDs derived both before (red squares) and after (black squares) pulsar gating. Also shown in the H.E.S.S. data of HESS J1023--575 (red open circles)
}
\label{fig:lc:sedpg}
\end{figure*}

The extended emission FGES~1023--5746  spatially coincides with the H.E.S.S. source HESS J1023--575. However, the low energy emission of this region is dominated by the pulsar PSR J1022--5746. To further investigate the emissions from FGES 1023--5746 we performed a pulsar gating analysis on PSR J1022--5746. We produced the phase-folded light curve of  PSR J1022--5746 by using   ephemeris  corresponding to the time period from MJD 54686 to 56583\footnote{see https://confluence.slac.stanford.edu/display/GLAMCOG/LAT+\\Gamma-ray+Pulsar+Timing+Models}. To obtain the light curve we adopted a $1^{\circ}$ aperture without applying any background subtraction. The resulted light curve is shown in Fig.~\ref{fig:lc:sedpg} (a). We note two peaks located in the phase 0.15 and 0.6, respectively. To reduce a possible bridge emission between the two peaks in the phase interval [0.2,0.5] we selected a rather narrow band [0.7,1.0] in the following analysis. We applied the same spectral  analysis described above. The derived SEDs are shown in Fig.~\ref{fig:lc:sedpg} (b), both for the emission before and after pulsar gating. We note that  the extrapolation of the low energy emission is significantly below the flux of HESS~J1023--575.  Yet, these two sources can be {\it indirectly} connected, namely if  the emission in the GeV band is dominated by the pulsar and the VHE emission is linked to the surrounding  pulsar wind nebula (PWN), as observed in many pulsar/PWN systems such as the Crab \citep{fermi_crab}. If we assume the pulsar emission has been eliminated in the pulsar gating process, the hardening and the extension are difficult to explain in a Pulsar/PWN scenario. However we cannot exclude the possibility that magnetospheric emission from the pulsar also exists in the bridge phase interval considered, dominating the GeV emission and preventing to derive an accurate of the extended emission coincident with the VHE source.  

\section{Gas around Westerlund 2}

The studies of the molecular hydrogen by  \citet{dame07}, \citet{furukawa09} and \citet{ohama10} show a  complex environment around Westerlund 2.  Two  molecular clouds in the velocity ranges of [-11 : 9] km/s and [11: 21] km/s  physically coincide with Westerlund 2 and and the large \ion{H}{ii} region RCW 49. The mass of the molecular gas is estimated to be $(1.7\pm 0.8)\times10^5 ~\rm M_{\odot}$.  The gamma-ray emission region, however, is much larger than Westerlund 2 and the corresponding molecular clouds. To study the molecular content in this region we  used data from the CO Galactic survey of  \citet{dame01}. We used the standard assumption of a linear relationship between the velocity-integrated 
CO intensity, $W_{\rm CO}$, and the column density of molecular hydrogen, N(H$_{2}$). The conversion factor $X_{\rm CO}$ is chosen to be $2.0\times10^{20} ~\rm cm^{-2} (K~km~s^{-1})^{-1}$ as suggested by \citet{dame01} and \citet{bolatto13}.  We  integrated the velocity interval between [-11: 21] ~\rm km/s. The derived molecular gas column density is shown in the left panel of Fig.~\ref{fig:gas} (a).

To determine the \ion{H}{ii} column density we used the {\it Planck} free-free map \citep{planck15-10}. We first converted the emission measure (EM) into free-free intensity by using the conversion factor in Table~1 of \citet{finkbeiner03}. Then we used Eq. (5) of \citet{sodroski97} to calculate the \ion{H}{ii} column density density from  the free-free intensity.  We note that the  \ion{H}{ii}  column density is inverse proportional to the electron densities $n_e$, which are chosen, following to \citet{sodroski97}, to be $2 ~\rm cm^{-3}$ and $10 ~\rm cm^{-3}$  as the upper and lower limits, respectively. The derived  \ion{H}{ii}  column density is also shown in Fig.~\ref{fig:gas} (b) (middle panel). To the southwest of RCW 49 there is another  bright \ion{H}{ii} region G287.393--00.630 \citep{anderson14}. However it has a radial velocity of $-17~\rm km/s$,  which is significantly different from that of RCW 49 ($\sim 0~\rm km/s$). G287.393--00.630 is likely to be a  foreground, although taking into account the uncertainties in the kinematic distance, we cannot exclude the possibility of physical correlation. To account for such uncertainties,  we estimated  two \ion{H}{ii} masses, one derived in the region where RCW 49 is located and another including both RCW 49 and G287.393--00.630. The total gas, adding the contribution traced by the CO and \ion{H}{ii} molecules, is shown is Fig.~\ref{fig:gas} (c) (right panel). 
The masses derived using different tracers are listed in Table \ref{tab:mass}. The total mass lies within  $2.0 \times 10^6 M_{\odot} < M <7.3 \times 10^6 M_{\odot}$. Assuming spherical geometry of the  \gray emission region,  its radius is estimated as $r = D ~\theta \sim 5.0 ~\rm kpc \times (2.4/57.29) ~\rm rad \sim 210 ~\rm pc  $. Thus, the averaged over the volume gas number density $n_{\rm gas}$ varies from  $7~\rm cm^{-3}$ to $25 ~\rm cm^{-3}$.

\begin{table}[h]
\footnotesize
\centering
\caption{Gas mass derived from different tracers. The lower and upper limits (in the region of RCW~49 or including also the region of G287.393--00.630) are indicated in the case of the free-free intensity tracer.} 
\label{tab:mass} 
\begin{tabular}{clclc}
\hline
\hline
Tracer &  Gas Phase & Mass ($10^6 ~\rm M_{\odot}$)\\
\hline
2.6 mm line & ~H$_2$ &~1.8\\
\hline
free-free intensity ($n_e = 2 ~\rm cm^{-3}$)   &  ~ \ion{H}{ii} & ~1.0 / 5.5 \\
\hline
free-free intensity ($n_e = 10 ~\rm cm^{-3}$)   &  ~ \ion{H}{ii} & ~0.2 / 1.1\\
\hline
\hline
\end{tabular}
\end{table}

\section{Origins of the extended \gray emissions}

Several scenarios can be put forward to explain the large \gray emission observed. Under the assumption of $\gamma$-ray production in an hadronic interactions of CRs interacting with the interstellar gas, the total \gray luminosity of $10^{36} ~\rm erg/s$ (for a fiducial distance of 5 kpc) and  gas density of $7 ~\rm cm^{-3}<n_{gas}< 25 ~\rm cm^{-3}$, result on a total CR energy confined in the $\gamma$-ray production area of $(4 - 13) \times 10^{49} ~\rm erg$. This estimate is  quite typical  for energy release in CRs by a SNR. Thus, at first glance, a single  SNR seems an attractive option as the source of production of CRs. However, below we argue that the injection of CRs into the interstellar medium seems to have a  continuous character giving a preference to the stellar cluster Westerlund 2 and the colliding wind binary Eta Carina as the suppliers of CRs.
\subsection {CRs continuously injected from Westerlund~2}

The young star cluster Westerlund 2 contains more than a dozen of  O stars and two Wolf-Rayet stars.  The total mechanical energy in the form of stellar winds  in this region is  estimated  around $10^{\rm 51}~\rm erg$ \citep{rauw07}, which could in principle provide a large energy reservoir to power the extensive \gray source \citep{parizot04}.  To account for this large emission, the acceleration of  particles, which escape their sources and interact with the surrounding  gas, should be accompanied by an efficient diffusion. Because of  large number of accelerators within the cluster, one could expect a
(quasi) continuous  injection of CRs into the interstellar medium, over the age of the cluster 
exceeding $T_0 \geq 10^6$ years. That would translate on a hard spectrum of \grays, mimicking the spectrum of accelerated protons, which are continuously accelerated and inflating the extended region.
Remarkably, a similar hard \gray emission has already been detected from two other young star associations:  Cygnus OB2 \citep{fermi_cygnus} and NGC 3603 \citep{yang17} with similar energy budgets in CRs ($1.3 \times 10^{49} ~\rm erg$ \citealt{fermi_cygnus} and $(2 - 10)\times 10^{49} ~\rm erg$ \citealt{yang17} respectively).     

We also note that it is possible that a significant part of CRs has already escaped from the \gray emission region. The relation between the total injected CR energy and the CR energy in the \gray emission region can be expressed as \citep{aa96}:

\begin{figure*}
\centering
\includegraphics[width=\linewidth]{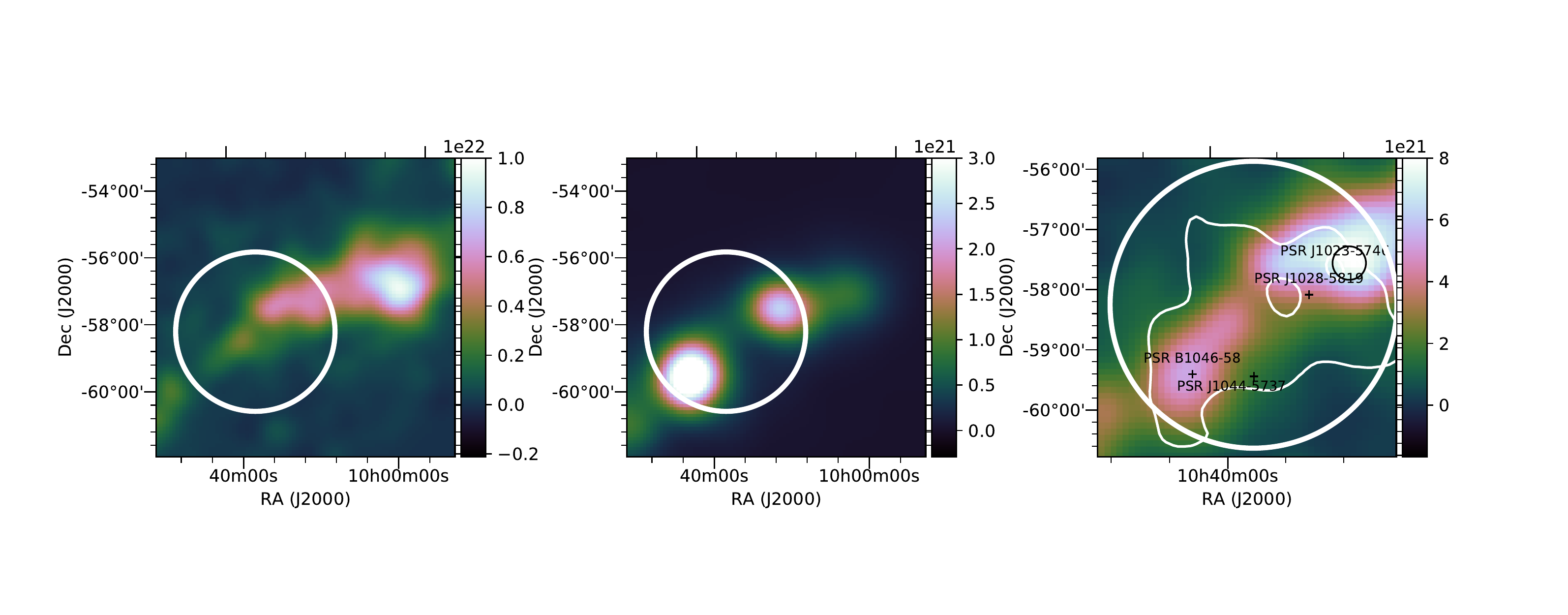}
\caption{Left panel a): Gas column densities derived from CO data, integrated in the velocity range of [-11, 21] km/s. Middle
panel b): \ion{H}{ii}  column density derived from the Planck free-free maps. Right panel c): Total gas column density by summing up the
molecular gas and \ion{H}{ii} gas. The \gray emission is overlaid with white contours together with the positions of the energetic pulsars in
the field and the position of Westerlund 2 (black circle). The white disk indicate the \gray template, in which the total gas mass is
calculated.}
\label{fig:gas}
\end{figure*}

\begin{equation}
\frac{n_{\rm emis}}{n_{\rm tot}}=(\frac{R}{r_{\rm diff}})^2=0.1\times (\frac{2\times10^6 ~\rm years}{T})(\frac{10^{28}~\rm cm^2/s}{D}), 
\end{equation}
where ${n_{\rm emis}}$ and ${n_{\rm tot}}$ are the CR density derived from the \gray luminosity and the total injected CR density, respectively. $R$ is the size of the \gray emission region and $r_{\rm diff}$  is the total diffusion length of  CRs corresponding to the age of the cluster T, and D is the diffusion coefficient.  In this case the total CRs energy is 10 times larger than that derived from \gray emissions and amount to $10^{51} ~\rm erg$. Such total CR energy can hardly be provide by a single SNR. Moreover, the shorter acceleration time (typically less than $10^4$ years  in the case of  SNRs) would result in a steeper spectrum due to the energy-dependent escape of CRs caused by diffusion.
In the case of a young star cluster such Westerlund 2, the wind mechanical energy of $10^{38} - 10^{39}~ \rm erg/s$ combined with the long injection time for an age of more than $10^6$ years, makes this an appealing scenario.

\subsection{Radial distribution of \gray emissivities } 

The spatial distributions  of CRs can provide key information of the injection history. The results obtained in this study do not show a strong correlation between the $\gamma$-ray and gas distributions which would be expected  in the case of homogeneously distributed CRs.  Meanwhile, in the case of continuous injection of CRs by the accelerators  located in Westerlund 2, into the surrounding extended regions beyond the cluster, we should expect  $1/R$ type radial  distribution of CRs, unless the diffusion coefficient  characterizing the propagation of CRs   suffer dramatic changes over 100-200  pc scales.

\begin{figure*}
\centering
\includegraphics[width=0.45\linewidth]{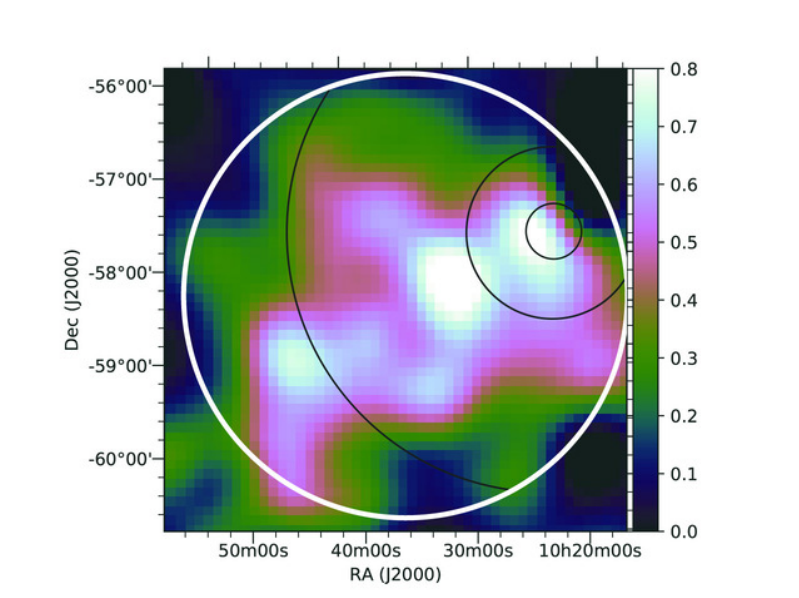}
\includegraphics[width=0.45\linewidth]{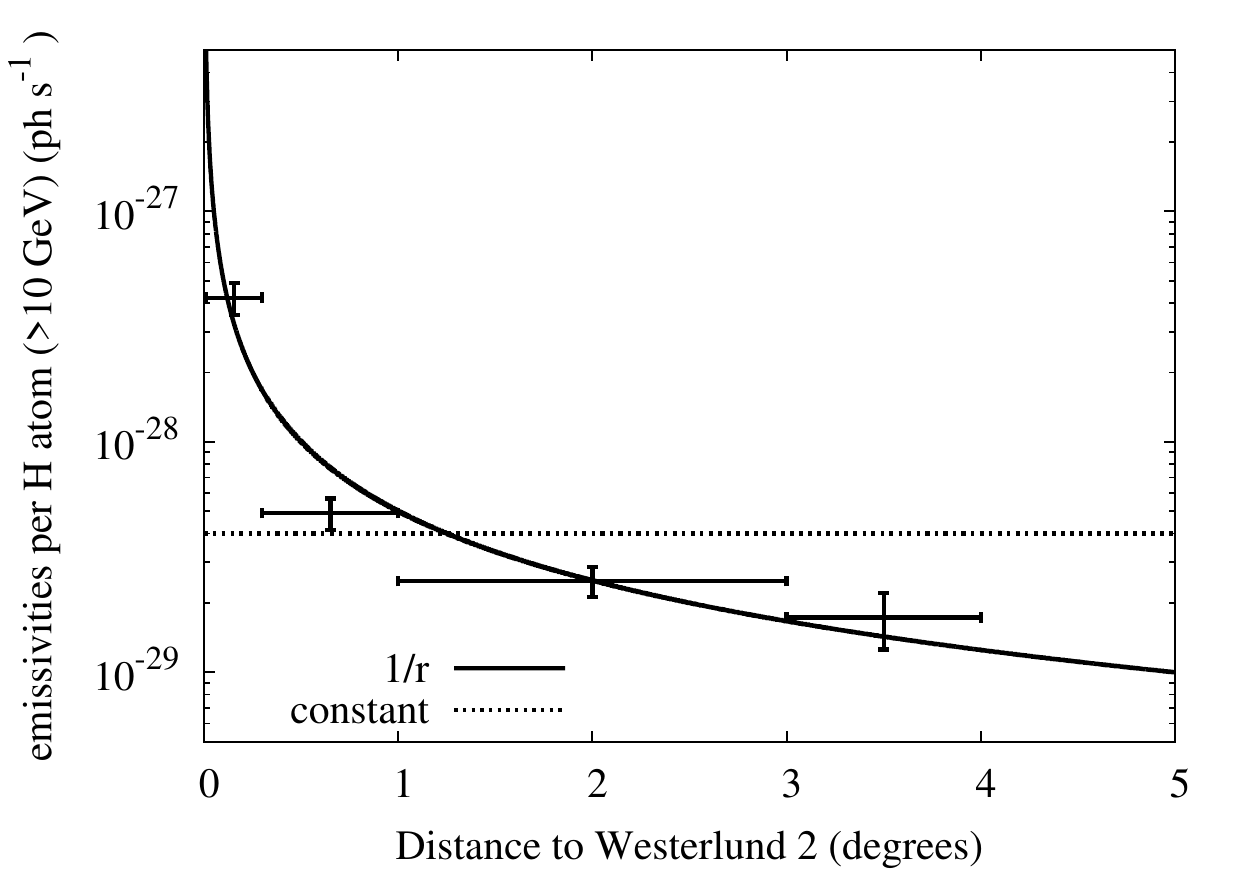}
\caption{{\it Left panel}: The white circle and black arcs label the regions to subtract the \gray emissivities. The position of Westerlund 2 is marked as cyan diamond.  {\it Right panel}: The profile of \gray emissivities per H atom above 1~GeV with respect to the distance to Westerlund 2. For comparison, we show $1/r$ (solid curve), and constant (dotted curve) profiles, which are predicted by the  continuous injection, and the impulsive injection, respectively.    }
\label{fig:pro}
\end{figure*}

To investigated the  CR distribution in the context of the the hadronic origin of the extended \grays emissions, we produce the radial profile of the the \gray emissivity per H atom, which is proportional to the CR density, taking the position of Westerlund~2 as reference. The \gray flux is derived above 10~GeV, using the standard likelihood analysis, in four regions resulting from splitting the 2.5$^{\circ}$ disk into subregions. They correspond to regions located to angular distances [0:0.3]$^{\circ}$, [0.3:1.0]$^{\circ}$, [1.0:3.0]$^{\circ}$ and [3.0:4.0]$^{\circ}$  away from Westerlund 2. The total gas column density is calculated for $n_e = 10 ~\rm cm^{-3}$ (which is the value suggested in \citealt{sodroski97} for the region inside the solar circle) and excluding the \ion{H}{ii} region G287.393--00.630. The resulting radial profile is shown in the right panel of Fig. \ref{fig:pro} (b).  We fit the density profile to two hypothesis: a constant value, which represent a homogeneous distribution of CRs formed in an impulsive injection event, and a  $1/r$ type distribution, which would result from a continuous injection of CRs from Westerlund 2. The latter is favoured with a $\chi^2$/ndf of 1.3 versus a $\chi^2$/ndf of 15.6 for the constant case, which strongly
support an scenario in which a central source (Westerlund2), is continuously injecting CRs into surrounding the region. 

 We want to stress here that the determination of CRs radial distribution  is a powerful tool to site localize the source and derive the injection history of  CRs. In particular, the 1/r distribution of CRs is a direct proof of the continuous injection of CRs. It can be regarded as a powerful method to identify the continuous CR sources in \gray astronomy.

The caveat of the results above is the uncertainty of the gas distribution. We  note the velocity integration range [-11, 	21] km/s derived from the compact clouds near Westerlund 2  could  be too high for the whole extended \gray emission region. \citet{furukawa09} have argued that the large velocity dispersion of molecular clouds near Westerlund 2 is caused by both stellar wind acceleration and the clouds collision.  The \ion{H}{ii}  column density are also uncertain due to the uncertainties of the  free electron density, although the  \ion{H}{ii}  column density has only minor influence on the results in this work.  On the other hand whether the \ion{H}{ii} region G287.393--00.630 are associated with the \gray emission is not known. A dedicated study on the ISM in this region is required  to pin down these uncertainties.  

\subsection{Other possible scenarios}

In principle, the diffuse \gray emissions can be dominated by the leptonic component
due to the inverse Compton (IC) scattering of electrons. Formally, a \gray photon  index 
close to 2  requires  relativistic electrons with the differential power-law index  of 3. This can be realised, for example, by the radiative cooling of electrons with an initial (acceleration) spectrum of $E^{-2}$. The IC component can  dominant in the GeV regime due to the presence of an enhanced target radiation fields (ISRF) supplied by the young stars of Westerlund~2. The total optical luminosity of all O and WR stars in  Westerlund 2 is estimated   as about $2 \times 10^{40} \rm erg/s$ \citep{rauw07}. For the size of the \gray emission region ($R \simeq 200$ pc), the contribution of these young stars to the ISRF is only $0.1 ~ \rm eV/cm^{-3}$, which is significantly lower than the diffuse starlight component in our Galaxy \citep[see, e.g., ][]{popescu17}.  More IC contribution is expected 
if the electrons are accelerated  near the stars, since they emit IC $\gamma$-rays 
in much denser radiation environment.  But in this case one would expect a rather compact \gray source limited by the angular size of Westerlund 2 cluster  contradicting the observations.

The colliding wind binary system Eta Carina is also located inside the \gray emission region. The powerful source has also
been detected as a compact source in the GeV and TeV regime with the LAT and H.E.S.S. telescopes \citep{FermiEtaCarina, HessEtaCarina}. While the high and VHE emission of the compact source may have either  leptonic or a hadronic origin, the extended diffuse $\gamma$-ray source should be dominated by interactions of  CR protons and nuclei with the ambient gas. The injection rate of these particles on timescales of more than $10^6$ years should be ,as in the case of Westerlund 2, as large as $10^{37}  ~\rm erg/ \geq 10~GeV$.

The extended emission could also be superposition of several discrete (unresolved) sources contributing to the total radiation. Indeed,  several pulsars and supernova remnants (SNRs) are located inside the \gray emission region.  The \gray spectra from  pulsars typically contain a cutoff at several GeV  thus  hardly can contribute to the $\geq 10$~GeV \grays detected in the extended structures.  At the same time, PWN and SNRs can be considered as natural contributors to hard \grays. 
Two SNRs, SNR G284.3--1.8  and G286.5--1.2  lie within the gamma-ray emission region. Both of them have a much smaller radio size compared to the size of the detected extended \gray emission. Considering the extension of the remnants, the extended emission can hardly be related to the SNRs themselves. TeV gamma-ray emissions has been detected towards the SNR G248.3--1.8, which is believed to be a combined contribution from the binary system 1FGL~J1018.6--5856 and the PWN associated to the bright pulsar PSR~J1016--5857 \citep{hess1018}. CRs escaping from those SNRs could also be appealed. In this case, the acceleration time lasts typically no more than $10^4$ years. Filling up a 200 pc region requires $2\times 10^5 (\frac{10^{28}~\rm cm^2/s}{D})^{0.5}~\rm years$. By then, the high energy CRs should already escape the region due to energy-dependent escaping and a soft spectrum would be expected, as observed in  $10^4 - 10^5 ~\rm years$ mid-age SNRs \citep[see, e.g., ][]{yuan12}.  

On the other hand, there are more than a dozen of pulsars located inside the \gray emission region. Four of them have high
spin-down luminosities, among which PSR\,J1023--5746 has a spin down luminosity of more than $10^{37}~\rm erg/s$.
Formally, this could be be sufficient to explain the detected \gray luminosity by high energy photons produced in the surrounding
pulsar nebulae. However the huge extension of the detected diffuse radiation (200 pc) significantly exceeds the typical size of
PWNe  \citep[see, e.g., ][]{gaensler06}. Thus a single PWN originating the whole
diffuse \gray emission seems also unlikely. The large emission
could still be due to the overlapping emission originated by PWNe
associated to those four bright pulsars pulsars (their positions indicated
in Fig. 1(b)): PSR\,J1023--574, PSR\,J1028--589, PSR\,J1044--5737 and PSR\,B1046--58. The first two are middle-age
and energetic (with spin-down power $\dot{E}$ of $1.1\times10^{37}$ erg/s and $8.3\times10^{35}$ erg/s respectively, \citealt{latpsr}) pulsars, for which the associated PWN has been detected in the TeV
regime \citep{hess_w2}. Despite the compact size of the TeV sources compare to the GeV ones, the bright northernmost GeV spots could
be related to a larger PWN emission due to the energy-dependent morphology \citep{HESSJ1825,HESSJ1303}. The last two pulsars have similar properties, with
high spin-down luminosities of $8.03\times10^{35}$  erg/s and $2.0\times10^{35}$ erg/s respectively, and could be powering similar PWNe to explain
the extension of the emission into high energies.

Observations with large field-of-view instruments with good angular resolution, are crucial to understand whether the emission is
powered by a single source or is a superposition of contributions from several  extended sources.

 \section{Conclusion}

The supper bubbles associated with OB associations/star-clusters has been 
proposed as possible production sites of the Galactic CRs  \citep{montmerle79, cesarsky83, bykov14}. One of the attractive features of this model is the explanation of  of the observed secondary CR abundances \citep{binns05,rauch09} . The recent observations of $^{60}Fe$ in CRs provide a new support to this hypothesis \citep{binns16}.  Furthermore, the measurements of \gray emissivities reveal a similar radial distribution of CRs with OB stars \citep{fermi_diffuse,yang16}.  If CRs are accelerated in such environments, high energy \grays are expected from the interactions of the quasi continuous injected  CRs with the ambient gas.  Here, we report a statistically significant detection of an extended \gray signal from the direction of the young star cluster Westerlund 2. The spectrum of this source is hard, and extends up to 250 GeV.  This detection, together with the similar reports 
regarding  Cygnus cocoon \citep{fermi_cygnus}, 30 Doradus C \citep{hess_lmc} , Westerlund 1 \citep{hess_w1, w1_lat} and NGC 3603 \citep{yang17},  may contribute to the better understanding of   the origin of galactic cosmic rays in general, and in the 
high energy phenomena in the super bubbles, in particular.

 In the context of the origin of galactic cosmic rays,  a special interest  is the 
 question  whether these objects can operate also as PeVatrons, i.e. whether they can provide the bulk of the locally observed CRs up to the so-called knee around 1 PeV.  
The most straightforward and unambiguous answer to this question would be the detection of  \grays extending with a hard energy spectrum to energies well beyond 10 TeV.  Apparently,  because of the limited detection area, the Fermi LAT observations cannot offer such measurements. In this regard the Imaging Atmospheric Cherenkov Telescopes (IACT) with their huge detection areas and adequate  angular and energy resolutions, are powerful tools for the search and study of cosmic PeVatrons.  Interestingly, multi TeV-\gray emission  with a hard spectrum already has been reported  from  Westerlund 2 by the HESS collaboration \citep{hess_w2old,hess_w2}. But the detected TeV emission has a much smaller extension than the one reported here.   The  limited field of view  of the current 
IACT arrays  is not optimal for  such extended structures. To deal with the large scale diffuse structures,  some  other data analysis algorithms such as the likelihood method seems to be  more  effective tool than the aperture photometry method used today in IACT data analysis. 
We should  note that a very extended TeV emissions has been reported from another similar young stellar cluster, Westerlund 1 \citep{hess_w1, w1_lat}. This result has been interpreted by \citet{bykov14} as an indication of acceleration of protons to PeV energies. However, this tentative result needs further confirmation.  A dedicated analysis on Fermi LAT data  taking advantage of the updated data calibration and the accumulated exposure also could be helpful.

\bibliographystyle{aa}
\bibliography{w2}
\end{document}